\documentclass{article}
\begin{document}
\hyphenation{re-pre-sen-ta-tion}
\hyphenation{va-ri-ous}
\hyphenation{va-ri-a-bi-le}
\hyphenation{con-si-de-ra-to}
\hyphenation{pro-dot-to}
\hyphenation{se-con-do}
\hyphenation{ex-pe-ri-men-ts}
\def\hb{{\hfill\break\indent}}
\def\hbn{{\hfill\break\noindent}}

\hbn
\centerline {\bf Error sensitivity of a Quantum Simulator I:}
\centerline {\bf a first example}
\hbn
\centerline { G. Strini${}^{*}$}
\hbn
\centerline {Dipartimento di Fisica dell'Universit\`a}
\hbn
\centerline {Via Celoria 16 Milano - Italy}
\hbn
\centerline {\bf Abstract}
\hbn
As the first useful Quantum Computers presumably will be quantum
simulators, here the minimum number of qubits necessary 
 for the solution of Schr\"odinger's equation in
 simple test problems
is evaluated.
From the present
preliminary results it appears that it is possible to realize a
useful quantum simulator with a register of only 10-15 qubits.
The intrinsic sensitivity to some errors 
appears to be moderate without the need of error correcting methods.
At present there is at least some indication that the amplitude errors
are more dangerous than phase and decoherence errors.
So corrections limited to amplitude errors may be very useful.

\hbn

\hbn
{\bf 1. Introduction}
\hbn

After the pioneering work of Shor on the 
factorization [1]  at present there is a growing
interest in the study of quantum computers (QC) and the related problems
of error control due to inaccuracies of the quantum gates,
decoherence and so on (for a still useful review see [2]).
As it is by now well known the QC may have an exponential advantage over the
classical ones on many  problems (for a very
incomplete list see [3, 4, 5, 6, 7]),
but due to implementation difficulties
the first QC  will  not have many qubits.
So it is almost obvious that the first useful QCs
will be quantum simulators and  not true numerical computers.

This is the first of a series of papers devoted to the
simulation of a quantum computer for the solution of a generic
quantum mechanical problem. The point of view is that of an
experimentalist wanting to know how many qubits are necessary
and what amount of errors it is possible to tolerate.
So the present problem is to find the minimum requirements
for the setting-up of a useful simulator to solve simple
problems.

The simulation of a quantum computer on a classical one was often done
in the past [8], but  for very specific problems.
Here we study a system of general use to
solve  standard undergraduate problems in quantum mechanics as
 can be found in an  exercise book [9].
Such a quantum simulator (QS) was simulated
on a classical 
 computer. Obviously the efficiency of such a simulation
is  exponentially slow, so nothing is gained in comparison with
 a direct solution of the various problems studied.
From the experience gained so far, it is fair to say that a simulation
of a QS on a classical computer approximately requires a
computational power similar to the one necessary for the direct
solution  on the classical computer, except for the 
simulation of some errors.
The simulation of the various kind of errors
is the most uncertain part of this work,  
 because at present there are not  sufficient experimental data
on the sources of errors in the various implementations of real systems.
It was then necessary to start with very simple hypotheses.
Various problems were simulated:
transmission and reflection of a Gaussian
wavepacket on various barriers, unidimensional 
harmonic and anharmonic
oscillators, the solution of the radial equation of the tridimensional
harmonic oscillator  etc..
Here we report only the results about the unidimensional
harmonic oscillator as the results of this simpler case 
are very similar to  those of the other
problems studied.

\hbn
{\bf 2. Solution of the Schr\"odinger equation}
\hbn

The algorithm we will use has been  described elsewhere [10, 11]
so let us merely show the essential points.
The Schr\"odinger equation:

\begin{equation}
i\, \hbar \, \frac{d}{d\,t}\, \psi (x;t) \;=\;
[-\, \frac{\hbar^{2}}{2\,m}\, \frac{d^{2}}{dx^{2}} \;+\; V(x)
] \, \psi (x;t)\; \equiv \;
[H_{0}\,+\, V(x)] \, \psi(x;t)
\end{equation}

\hbn
may be easily solved on a QC provided that it is possible to decompose the 
algorithm in a sequence of unitary operations
enough simple and, what is more important, efficiently implementables
on a simple quantum computer.

The first step is to decompose the region of interest
of the space coordinate in $2^{n}$ intervals of length $\Delta$ and to
represent these intervals in the  Hilbert space of a $n$
qubits register. This space is the direct product of $n$ copies of the 
space of a single qubit:

\begin{equation}
|i_{n-1} \rangle \,\otimes\,
|i_{n-2} \rangle \,\otimes\,
\cdot
\cdot
\cdot
\,\otimes\,
|i_{0} \rangle
\;=\; | i \rangle
\end{equation}

\hbn
where $(i_{n-1}, i_{n-2},..., i_{0})$ is the binary decomposition
of the index $i$.

Then, neglecting  the simbol $\otimes$, 
 the wavefunction $\psi(x;t)$ is represented by the superposition:

$$
\psi(x;t) \;\equiv\;
 \sum_{i=0}^{2^{n}-1} \, C_{i}(t) \, |i \rangle
\;=\;
\sum_{i_{n-1}=0}^{1}\,
\sum_{i_{n-2}=0}^{1}\,
\cdot
\cdot
\cdot
\sum_{i_{0}=0}^{1}\, C_{i_{n-1},i_{n-2},...,i_{0}}(t)\,
|i_{n-1}\rangle
|i_{n-2}\rangle
\cdot
\cdot
\cdot
|i_{0}\rangle
\;=\;
$$

\begin{equation}
 \;=\; \frac{1}{N}\,
\sum_{i=0}^{2^{n}-1}\, \psi (x_{i} \,=\,
 (i + 0.5)\, \Delta ;t)
|i\rangle 
\end{equation} 
\hbn 
where $N$ is a normalization factor as the samples are not normalized.
From this representation it is evident the exponential advantage
to use a QS instead of a classical computer: a register
of $n$ qubits can store $2^{n} - 1$ complex numbers.
 This exponential advantage
is completely lost in the simulation of the QS on a classical
computer.

To implement the various operations to solve the (1) it is necessary to 
extend the Hilbert space of the QS given by the register
$|n\rangle$ with an ancella register  $|m\rangle_{a}$
of $m$ qubits. So the
complete Hilbert space of the QS  now reads:

\begin{equation}
|m \rangle_{a} \, \otimes \, |n\rangle
\end{equation}

For the simulation of some errors of this QS on
a classical computer it is necessary to add in the classical simulator the
simulation of other qubits $|l\rangle_{l}$ to describe unwanted
quantum degrees  of freedom or a rough representation of the
environment.
The complete Hilbert space of the classical simulation of the QS thus is:

\begin{equation}
|l\rangle_{l}\,\otimes\,
|m\rangle_{a}\,\otimes\,
|n\rangle
\end{equation}

\hbn
and for the classical simulator it is necessary to store 
 $2^{(l+m+n)}$ complex numbers.
The experience gained shows that the practical 
 limits of the simulation appears to be due more to
 problems of storage than  to computational power.

The Eq.(1) may be formally integrated by propagating the initial wavefunction 
every timestep $\epsilon$:

\begin{equation}
\psi\, (x;t \,+\, \epsilon) \;=\; 
e^{-\, \frac{i}{\hbar} \, [H_{0} \,+\, V(x)]
 \, \epsilon} \, \cdot \, \psi\, (x;t)
\end{equation}

Using a time interval $\epsilon$ 
small enough to  neglect terms
in $\epsilon^{2}$
 it is possible to write:

\begin{equation}
e^{-\, \frac{i}{\hbar} \, [H_{0}\,+\, V(x)] \, \epsilon}
\;\approx\;e^{-\, \frac{i}{\hbar} \, H_{0} \, \epsilon}\,\cdot\,
e^{-\, \frac{i}{\hbar} \, V(x) \, \epsilon}
\end{equation}

These operators are still unitary,  simpler than the initial one 
and  efficiently implementable on a QC.
As the Fourier transform ($\mathcal{F}$) is also easily implemented [2]:
by calling  $k$  the transformed variable, it is possible to write  the
first operator in the right hand side of (7) as:

\begin{equation}
e^{+\, \frac{i}{\hbar} \, (\frac{\hbar^{2}}{2\,m
}\, k^{2}) \, \epsilon} \;=\; U
\end{equation}
\hbn
and so in the space representation:

\begin{equation}
e^{- \, \frac{i}{\hbar} \, H_{0} \, \epsilon}\;\equiv \;
{\mathcal F}^{-1} \, \cdot\, U \, \cdot \, {\mathcal F}
\end{equation}

The problem is now reduced to the implementation of an
operator of the form:

\begin{equation}
e^{i \, f(x)} \, |x\rangle
\end{equation}

This is possible
using an ancella quantum register $|m\rangle_{a}$
 by means of the following steps:

\begin{equation}
 |0\rangle_{a} \, \otimes \, |x\rangle
\rightarrow
 |f(x)\rangle_{a} \, \otimes \, |x\rangle
\rightarrow
e^{i \, f(x)} \, 
|f(x)\rangle_{a} \, \otimes \, |x\rangle
\rightarrow
\end{equation}

\begin{equation}
\rightarrow
e^{i \, f(x)} \, 
|0\rangle_{a} \, \otimes \, |x\rangle
\equiv
|0\rangle_{a} \, \otimes \, e^{i \, f(x)} \, 
|x\rangle
\end{equation}

The first step is a standard binary representation of the function 
$f(x)$ on the ancella register and may be  implemented by means
of  c-not gates.
The second step is the operation $e^{i\,m}|m\rangle_{a}$
done on the ancella qubits and so efficiently implemented
by means of standard techniques with  phase gates giving a phase shift 
to every qubit proportional to the weight of the  qubit.
The third step is just the reverse of the first one and
immediately implemented by the same array of gates as the
first one but applied in the reversed order.
It may be useful to note that, in the examples studied,
the implementation required 
hundreds of c-not gates.
As the ancella qubits are returned to the standard configuration
it is possible to use the same ancella qubits for the
kinetic energy and the  potential energy computations.
The potential energy part of the (7) is implemented in a
similar way.
In conclusion this algorithm
  can be implemented
with no problems, except
 for some tricks 
necessary to minimize the number of the space and ancella qubits.

\hbn
{\bf 3. Dimension of the  registers}
\hbn

As long as at present qubits are a very expensive resource,
the first important problem to be  solved 
is to determine  the minimum number of qubits.
Here we devote a special care to this problem as it is usually
totally neglected.

For the computation are needed
 $n$ qubits to represent $2^{n}$ points of 
the coordinate space and $m$
  ancellae qubits are also needed for the computation of
the potential and the kinetic energy
with the resolution of $2^{m}$ steps.
The most simple determination is the number $n$ of the
$2^{n}$ space points. This 
problem is common to every  numerical 
representation of functions and is solved
in a well known way by referring to the sampling theorem.
As this theorem [12] is of central importance
 to keep the number of the qubits  as small
as possible it may be useful to remember some details. 
This theorem allows to compute the errors in reconstructing 
a  function from its samples. 	
It is enough to remember that the
 sampled function may be represented as the product
of the original function times a train of Dirac delta functions.
As long as the transform of a train of Dirac deltas is still a train 
of Dirac deltas,
one gets  replicas of the Fourier transform of the
original function, shifted in frequency  by the interval
determined by the sampling frequency. 
So for
a function whose spectrum is limited to a maximum frequency it is
enough to sample at a frequency at least the double of this maximum
frequency  to avoid the superposition of the tails
of the spectrum.
At a lower sampling frequency there is an error in  reconstruction due to 
 tails superposition. This kind of error is  referred to
as aliasing error.
To be concrete in the present problem it was assumed an initial
wavefunction with a Gaussian shape. Since
  the Fourier transform
 is still a Gaussian 
so that the spectrum goes to zero very rapidly  a
 surprisingly small number of samples was sufficient helping to
keep the space register small.

The implementation of functions is relatively simple: it is enough
to "write" the values of the function in the binary representation
in an ancella register of $m$ qubits  with a
resolution of $2^{m}$ steps. This writing is realized
by means of an array of c-not gates (the details will be
given elsewhere).
A hard task  is the evaluation of 
the number
$m$ of the ancella
qubits necessary for the potential and kinetic energy computations.
For this problem
unfortunately there is not a theorem similar to the sampling theorem 
but only a less useful one on the characteristic function of
the probability density of the
function to be digitized (for a tutorial on these problems see [13]
and cited references).
So it was necessary to use some numerical tests.

An unexpected result  was that usually the number of
the ancella qubits outweights the number of the qubits representing the
space coordinate.
The most sensitive part of this problem was the computation of the kinetic
energy after the  Fourier transform of the space wavefunction.

A first trick
to keep this number as small as possible is to use wavefunctions  smooth
in the space coordinates, in order   not to  have strong
components of short wavelength.
To have a concrete idea of what is happening, in
problems with a  simple potential as the harmonic oscillator
in the case of an initial Gaussian wavepacket with a spread about 1/50 of the
range of the space coordinate, and using only $n=6$ qubits, 
i.e. $2^{6}=64$ points
for the space coordinate,
 the components of the Fourier
transform of the wavefunction
are populated only in the very lower part of the spectrum. 
In this case,
 for the computation of the kinetic energy up to $m=12$
ancellae qubits were necessary to reach a
sufficient resolution.
A second trick  to keep this number small was to
compute by means of the ancellae the kinetic energy 
limited to the most populated part of the spectrum
and to  disregard the rest, hence using the full resolution of $2^{m}$
steps only for the used part of the spectrum.
Tipical values used after various tests, were $n=6$
space qubits and $m=7$ ancella qubits.
A total of only 13 qubits was sufficient for most tests.

\hbn
{\bf 4. Sensitivity to memory errors}
\hbn

A further  problem is the evaluation of the
sensitivity of the simulator to errors.
In the past various studies of the sensitivity to errors
were done on
numerical QCs [14, 15, 16, 17] and quantum computers
dedicated to special problems [7], but here we
concentrate on simulators
able to solve a generic quantum mechanical problem.
It is necessary  to define exactly which kind of errors are
to be studied and then to give an evaluation of the effect of these errors.
The effect of the errors  depends
on the observable  of interest. Here
the effect was estimated by computing the fidelity
 defined by the comparison of the erroneous
wavefunction 
$|\psi\rangle_{err}$ with the correct one
$|\psi\rangle$ computed using the same system and zero errors.

\begin{equation}
 f \;=\; |\langle \psi | \psi \rangle_{err}|^{2}
\end{equation}

This quantity gives the probability that  the erroneous wavefunction
will  pass a test for the correct one.
The first kind of errors studied are the memory errors.
These errors are simulated by applying a generic unitary
matrix with random parameters to  one qubit at a time 
 of the register
representing the space coordinate.
 This matrix:

\begin{equation}
U\;=\;
\left[
\begin{array}{cccc}
cos\theta & exp\{ i \, \alpha\} \cdot sin\theta \\
-\,exp\{ i \, \beta \} \cdot sin\theta & exp\{ i \, ( \alpha \,+\, \beta)
\cdot cos\theta\}
\end{array}
\right]
\end{equation}

\hbn
is applied at every timestep with random parameters
with uniform distribution between plus and minus a given maximum 
 (only a single parameter at a time is
different from zero:
as  the errors are assumed to be small the probability of two or
more simultaneous errors is taken to be negligible).
To ease the comparison 
of the effect of
different errors the same string
of random numbers was used in all the tests.
To be more precise, the integration of  Schr\"odinger's equation
was done for 40 timesteps and the error was applied
every step
 between steps
10 and 30.

The dependence of the fidelity 
 on the error parameter, the
maximum  amplitude of the
error and the qubit in error have been explored.
In the parametrization used the parameters $\alpha$ and $\beta$
have an analogous effect so it is enough to study only the
effect of the  errors described by
 the $\alpha$  and $\theta$ parameters.
For brevity we refer as $\alpha,\,\beta$ and $\theta$ errors
the errors due to the $U$ matrix with only the specified
parameter different from zero.
Also for error $\theta=0.1$ radiants we mean a series of $U$
matrices with only the parameter $\theta$ different from zero and
random with uniform distribution beetwen $\pm 0.1$ radiants. 
Similarly for the other parameters.
The results show 
a moderate effect of errors  $\alpha$ (and so also  $\beta$),
and a strong effect of the $\theta$ errors.

The errors are applied from the steps 10 to 30 of the 40 integration, 
so that the fidelity starts with the value 1, remains in this value
until the first application of the error matrix and then decreases
randomly to the minimum when the last error is applied. 
In Table 1 we report this final value of the fidelity resulting
from the error matrix applied to the various qubits of the
register representing the space coordinate.

\hbn

\centerline {Table 1. Final fidelity from $\alpha$ and $\theta$ errors}

$$
\begin{array}{cccccc}
qubit      & \alpha = 0.30   &
\theta = 
 0.05  & 
\theta = 
0.10   &
\theta = 
 0.20 &
\theta = 
 0.30 \\
  &  &  &  &  & \\
0  & 0.90    & 0.99  & 0.96   & 0.83 & 0.66  \\
1  & 0.90   & 0.99  & 0.96   & 0.85 & 0.70  \\
2  & 0.90   & 0.99  & 0.97   & 0.89 & 0.76  \\
3  & 0.83   & 0.99  & 0.96   & 0.86 & 0.71  \\
4  & 0.96   & 0.98  & 0.91   & 0.69 & 0.43  \\
5  & 0.98   & 0.98  & 0.93   & 0.75 & 0.53
 \end{array}
$$

The first column indicates the qubit acted-on by the error matrix (14).
The second column labeled  $(\alpha=0.30)$
gives the final
fidelity obtained with errors of maximum $0.30$ radiants on the parameter
$\alpha$ of the matrix (14). The following 4 columns give
the final 
fidelity for errors with the parameter $\theta$ respectively of maximum 
0.05, 0.10, 0.20 and 0.30 radiants.
By comparing the second and the last column it appears that
the average effect of the errors in parameter $\alpha$ are almost
 negligible in comparison with the error in parameter $\theta$.
It is possible to note an unexpected low 
 sensitivity to $\theta$
errors as high as 0.10 radiants, as shown in the $4^{st}$ column.
Another unexpected result is the low correlation of the
fidelity with the number $n$ of the qubit in error
(of weight $2^{n}$)
 indicated in the
first column.

The figures 1 and 2 give a plot of $|\psi(x;t)|^{2}$ for $\theta$
errors in the qubits "0" and "5". It is useful to give a physical 
interpretation of these errors. This error gives a probability
of reversal of the value of the qubit acted-on by the error.
Upon indicating with  $\Delta$ the step
in the space coordinate, the $\theta$ error in the qubit "0"
gives a reshuffling of the wavefunction values in adiacent intervals.
A $\theta$ error in the qubit "1" gives an analogous reshuffling
for intervals $2\Delta$, and so on for the $\theta$ errors
in the other qubits of higher weight.

As the wavefunction used is a Gaussian wavepacket there are at least two 
different regimes depending on the width of the reshuffling interval,
namely if it is smaller or greather than the width of the
wavepacket. This effect is clearly evident in figure 1 and 2,
referring respectively to $\theta$ errors in the qubits "0" and "5"
and so referring to space intervals of lenght 
$\Delta$ and $32 \Delta$ respectively.


\hbn
{\bf 5. Effect of the leak errors}
\hbn

Other sources of errors are  quantum leaks, gate errors and decoherence.
Here we limit ourselves to leak errors.
The quantum leaks are due to  unwanted interactions of the qubits of the 
working register of the QS with
quantum degrees of freedom
not involved in the computation.
To simulate these errors it is necessary to add some auxiliary qubits
$|l\rangle_{l}$ to the simulator Hilbert space. So this space now
is given by (5).

Here we study the effect of the leak errors due to the interaction
of a single qubit of the space register with a single extra quantum degree
of freedom described by a single qubit starting in a pure state.
So it is possible to represent this interaction by means of a unitary
4x4 matrix applied to a two qubit space given by 
the qubit supposed to be acted by the error and the
leak qubit at the specified times and tracing away the ancella leak
qubit at the end of the computation.
Also in this case the error matrix is applied every timestep
between steps 10 and 30.
The error matrix is now 4x4 and so it is in general described by
too many parameters to be useful. So 
in the absence of sufficient experimental data
we did  select two particular
very interesting matrices $U_{1}$ and $U_{2}$ 
depending from a single parameter 
given by [18]:

\begin{equation}
U_{1} \;=\;\left[
\begin{array}{cccc}
1  &      0      & 0 & 0 \\
0  & cos\theta & 0 & -\, sin\theta \\
0  & 0 & 1 & 0 \\
0  &  sin\theta & 0 & cos\theta
\end{array}
\right]
\hspace{0.5 in}
U_{2} \;=\; 
\left[
\begin{array}{cccc}
1  &      0      & 0 & 0 \\
0  & cos\theta & 0 & sin\theta \\
0  & sin\theta & 0 & - \,cos\theta \\
0  & 0 & 1 & 0
\end{array}
\right]
\end{equation}

To have an idea of the reason of the choice of these matrices, it is
useful to note that the matrix $U_{1}$, when applied to
a system composed of a single qubit and a single ancella qubit gives
a pure
decoherence. In fact denoting with  $X,Y,Z$ the parameters of the Poincar\'e
sphere of the density matrix of a single qubit:

\begin{equation}
\rho \;=\; \frac{1}{2}\,
\left[
\begin{array}{cc}
1 \,+\, Z  &  X \,-\, i\, Y \\
X \,+\, i \, Y  & 1 \,-\, Z
\end{array}
\right]
\end{equation}

\hbn
the  matrix $U_{1}$ 
applied to a system composed of a single qubit with this
density matrix and an ancella qubit in the pure state $|0\rangle$ 
after the trace
of the ancella qubit gives for the system 
reduced density matrix the transformation:

\begin{equation}
X' = cos\theta \,X
\hspace{.5 in}
Y' = cos\theta \,Y
\hspace{.5 in}
Z' = Z
\end{equation}

\hbn
and so the coherence $(X-iY)$ is reduced.
The
analogous computation with the
 matrix $U_{2}$ gives:

\begin{equation}
X' = cos\theta \,X
\hspace{.5 in}
Y' = cos\theta \, Y
\hspace{.5 in}
Z' = sin^{2}\theta \,+\,
cos^{2}\theta \, Z
\end{equation}

\hbn
and so the coherence is reduced and there is also a shift
and a factor lower than 1 of the $Z$
component of the reduced density matrix representing the population.
The fidelity 
resulting from errors due to the application of the 
$U_{1}$ and the $U_{2}$ matrices to the various qubits of the
space register 
was computed in a  way similar to the preceeding
case but with the  reduced matrix obtained by tracing away the ancella 
qubit $|l\rangle_{l}$.

Similar to the previous case, the fidelity starts with the value 1 and then
randomly decreases to the final value when the last error matrix
is applied. The parameter $\theta$ is supposed to be random
with uniform distribution between plus and minus the maximum value indicated
and the error matrices are applied from the timestep 10 to
the timestep 30.

\hbn

\hbn
\centerline { Table 2. Final fidelity from leak errors   $U_{1}$}

$$
\begin{array}{cccc}
qubit  &  
\theta = 
0.05 &
\theta = 
  0.10 &
\theta = 
  0.30  \\
 & & & \\
0  &  1.0  & 0.99   & 0.82    \\
1  &  1.0  & 0.98   & 0.89   \\
2  &  1.0  & 0.99   & 0.79   \\
3  &  1.0  & 0.99   & 0.77   \\
4  &  1.0   &  1.0  & 0.95   \\
5  &  1.0   &  1.0  & 0.93 
  \end{array}
$$

\hbn

\hbn
\centerline { Table 3. Final fidelity from leak errors   $U_{2}$}

$$
\begin{array}{cccc}
qubit & \theta = 
0.05 &
\theta = 
  0.10 &
\theta = 
  0.30  
\\
 &  &  &  \\
0 &    1.0  & 0.98   & 0.76    \\
1  &
    1.0  & 0.99   & 0.83   \\
2 
  &  1.0  & 0.97   & 0.82   \\
3
  &  1.0  & 0.94   & 0.78   \\
4
  &  0.99   &  0.97  & 0.79   \\
5
  &  1.0   &  0.97  & 0.66
 \end{array}
$$

In Table 2 we report the final
fidelity for leak errors on the various qubits
of the register of space coordinate.
The first column indicates the qubit acted-on by the leak.
The following three columns give the final 
 fidelity for various values of the maximum of
the error parameter $\theta$ in the matrix $U_{1}$.
The Table 3 give the analogous quantity for the matrix $U_{2}$.
In both cases
there is a low correlation
of the fidelity with the weight of the qubit in error.

The leak error is due to the transfer of
some of the quantum information on the system to a further
 quantum degree of
freedom. But as the 
computation of the fidelity
 is done by neglecting this degree
of freedom
and  calculating the reduced density matrix,
 part of the quantum information is regained.
This effect may explain the (lower than expected) effect of the quantum leak,
even with an error as high as the 0.30 radiants of the 
parameter used in the matrices $U_{1}$ and $U_{2}$.

In figure 3 and 4 we report a plot of $|\psi(x;t)|^{2}$ for the
errors due to the matrix $U_{1}$ applied to the space qubit "0"
and "5" for the error $\theta=0.1$ radiants, corresponding to the
third column of Table 2.
It is possible to note a very small effect of this kind of
error.
In figure 5 and 6 are reported the analogous results for the matrix
$U_{2}$. 
 The matrix $U_{2}$ generates, along with
decoherence errors, also population errors.
In this case it is possible to distinguish at least
two different regimes.
 This last kind of error
gives an effect similar to the $\theta$ error from the matrix $U$
shown in Table 1 and figure 1 and 2.
As in the case of the figure 1, in figure 5 it is evident a granularity
due to the reshuffling of the wavefunction values
 in adjacent space intervals.
This granularity changes to a random noise in figure 6 referring
to errors in qubit "5" as in the corresponding $U$ case
reported in figure 2.

By comparing  the results obtained
from the matrix $U_{1}$ (decoherence) and those
from the matrix $U_{2}$ (decoherence and population), it appears that
the last kind of errors is more dangerous.

\hbn
{\bf 6. Conclusions and future developments}
\hbn

The first  aim of this work was to have reliable data on how many
qubits are necessary to implement a QS for the solution of a generic and
simple quantum mechanical problem.
Susprisingly the number of ancella qubits necessary for the implementation
of the various operators outweights
the number of 
 the qubits representing
the space  degree of freedom of the system. But for the simple
problems studied here, with some care a total of only 13 qubits 
proved to be sufficient.
Another useful
 result is that the QS appears to be quite robust against
the explored errors: an error of 0.1 radiants on all the parameters
indicated gives a fidelity higher than about 0.9. 
Obviously for other observables and for a different measure
of the error this conclusion
may be not valid.
The fidelity is largely independent
from the weight of the qubit in error.

To ease the comparison of the memory errors as defined here
with other definitions, it is sufficient to say that the $\alpha$
and  $\beta$
 errors are similar to the phase errors and the $\theta$ error is 
similar to the amplitude error.
So it is possible to say that in this QS the amplitude errors
are more dangerous than the phase errors.

The most important development is to enlarge the single leak qubit to a
larger space in order to model the environment.
The interaction of a single qubit with the 
environment  has been modeled by various authors[19, 20]; 
it is known that two ancella qubits are sufficient,
but one needs to trace away these qubits at every integration
step in order to wash-out any memory of the system taken by the
environment.
Hence it is necessary to describe the system by means of the density matrix.
This does not create any basic problem, but only problems of storage
and computational power of the classical computer.

\hbn
{\bf Acknowledgments}
\hbn

This work was supported by M.P.I.
I would like to thank prof. V.G. Benza for his careful reading
of the manuscript and the many useful suggestions.

\hbn

(*) E-mail giuliano.strini@mi.infn.it

\hbn

{\bf References}
\hbn
[1] P. W. Shor. Proc. 35th Ann. Symp. Found. Comp. Sci.
{\bf IEEE Comp. Soc. Pr.} 124 (1994).

\hbn
[2] A. Eckert and R. Jozsa. Reviews of Modern Physics {\bf 68},  733 (1996).

\hbn
[3] S. Lloyd. Science {\bf 273}, 1073 (1996).

\hbn
[4] D. S. Abrams and S. Lloyd. Phys. Rev. Lett. {\bf 79}, 2586 (1997).

\hbn
[5] D. A. Lidar and H. Wang. Phys. Rev. {\bf E59} 2429 (1999).

\hbn
[6] D. A. Lidar and O. Biham. Phys. Rev. {\bf E56} 3661 (1997).

\hbn
[7] G. Benenti, G. Casati, S. Montangero and
D. L. Shepelyansky. arXiv:quant-ph/0107036.

\hbn
[8] J. Wallace Quantum Computer Simulators. A Review. 
http://www.dcs.ex.ac.uk/~jwallace.

\hbn
[9] I. I. Gol'dman et al.. Problems in Quantum Mechanics. 
Infosearch Limited. London 1960.

\hbn
[10] C. Zalka. Fortschr.   Phys. {\bf 46}, 877 (1988).

\hbn
[11] S. Wiesner. arXiv:quant-ph/9603028.

\hbn
[12] F. M. Reza. An introduction to information theory. McGraw-Hill 1961.

\hbn
[13] A. Grassi and G. Strini. Alta frequenza  {\bf  XXXIII}  E547 (1964).

\hbn
[14] K. M. Obenland and A. M. Despain. Proc. of High-Performance
Computing 
{\bf HPC'98} 228 (1998).

\hbn
[15] C. Miquel, J. P. Paz and R. Perazzo. Phys. Rev. {\bf A54} 2605 (1996).

\hbn
[16] A. Barenco and A. Ekert. Phys. Rev. {\bf A54} 139 (1996).

\hbn
[17] C. Miquel, J. P. Paz and W. H. Zurek. 
Phys. Rev. Lett. {\bf 78} 3971 (1997).

\hbn
[18] Chi-Sheng Niu and R. Griffiths. Phys. Rev. {\bf A60}, 2764 (1999).

\hbn
[19] B. Schumacher. Phys. Rev. {\bf A54}, 2614 (1966).

\hbn
[20] G. Strini. Unpublished Lecture Notes (2000).
\hbn
\hbn
\hbn
\newpage

\centerline {Figure captions}
\hbn

Figure 1. Plot of $|\psi(x;t)|^{2}$ obtained with the memory error matrix
$U$ applied to the qubit "0" of weight $\Delta$. The 
granulosity due to the reshuffling
 of the wavefunction values at a distance $\Delta$ is clearly
visible.

\hbn

Figure 2. Plot of $|\psi(x;t)|^{2}$ obtained with the memory error matrix
$U$ applied to the qubit "5" of weight $32 \Delta$. The reshuffling
 of the wavefunction values at a distance $32\Delta$
originates a random noise.

\hbn

Figure 3. Plot of $|\psi(x;t)|^{2}$ obtained with the leak error matrix
$U_{1}$ applied to the qubit "0" of weight $\Delta$. This leak 
originates a decoherence error and a very low noise.

\hbn

Figure 4. Plot of $|\psi(x;t)|^{2}$ obtained with the leak error matrix
$U_{1}$ applied to the qubit "5" of weight $32 \Delta$.
As in the preceeding figure 3 this leak originates
a decoherence error and a very low noise.

\hbn

Figure 5. Plot of $|\psi(x;t)|^{2}$ obtained with the leak error matrix
$U_{2}$ applied to the qubit "0" of weight $\Delta$. This leak 
originates a decoherence error and an amplitude error. The amplitude error
in the qubit "0" originates a reshuffling of the 
  wavefunction values at a distance $\Delta$ and so a
granulosity very similar to the one
due to memory errors reported in figure 1.

\hbn

Figure 6. Plot of $|\psi(x;t)|^{2}$ obtained with the leak error matrix
$U_{2}$ applied to the qubit "5" of weight $32 \Delta$. This leak 
originates a decoherence error and an amplitude error. The amplitude error
in the qubit "5" originates a reshuffling of the 
  wavefunction values at a distance $32 \Delta$
and so 
originates a random noise similar to the one
due to memory errors reported in figure 2.

\hbn

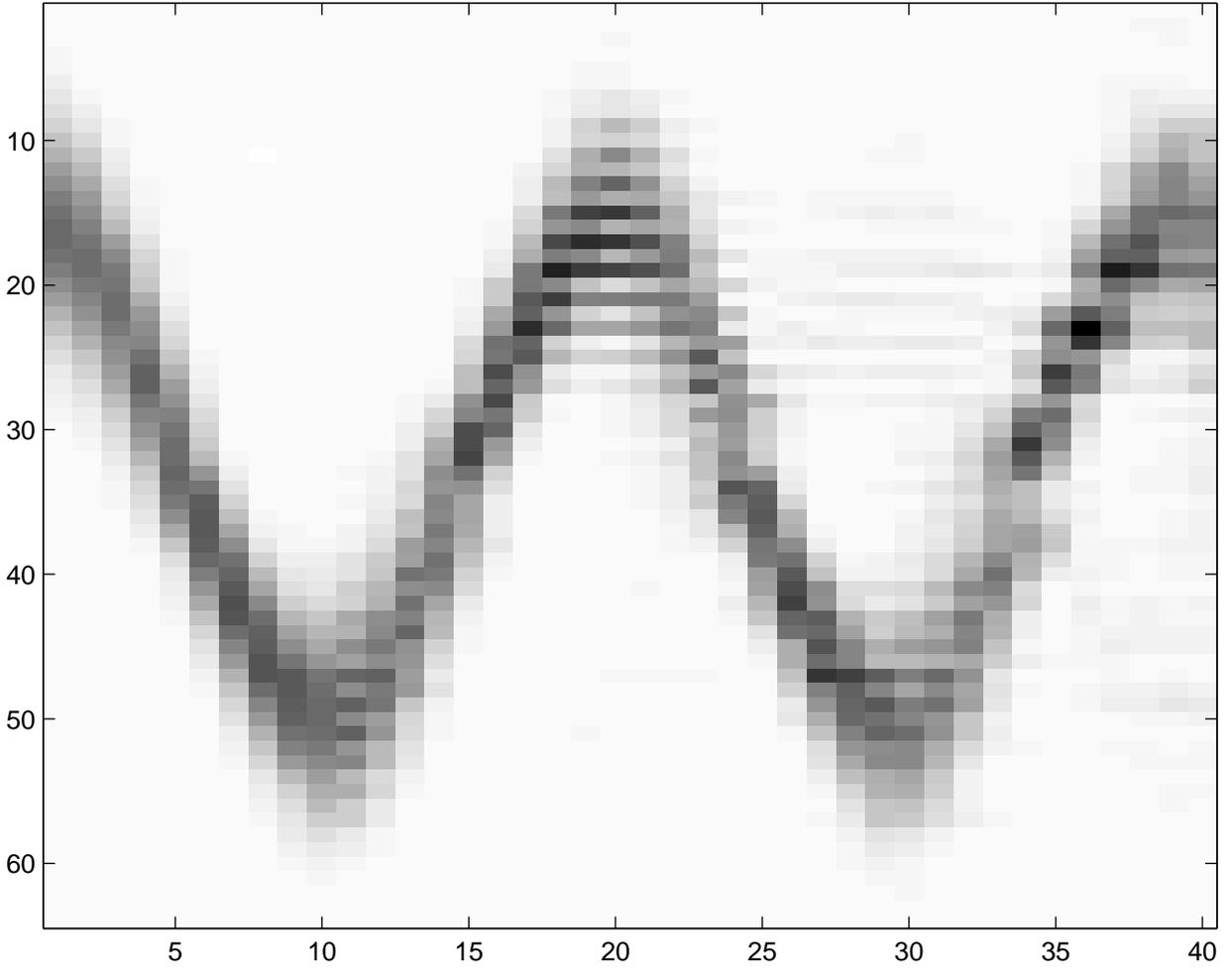
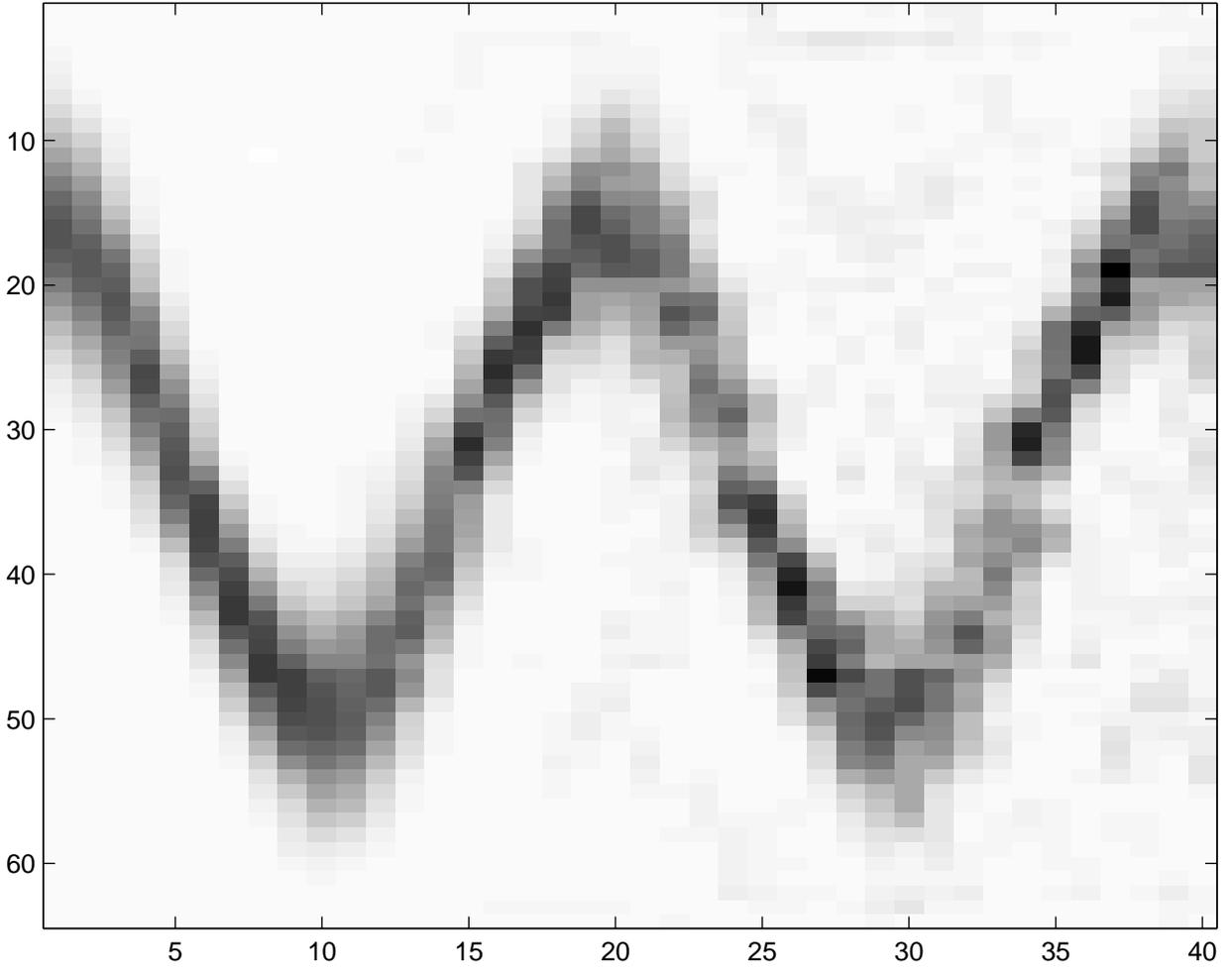
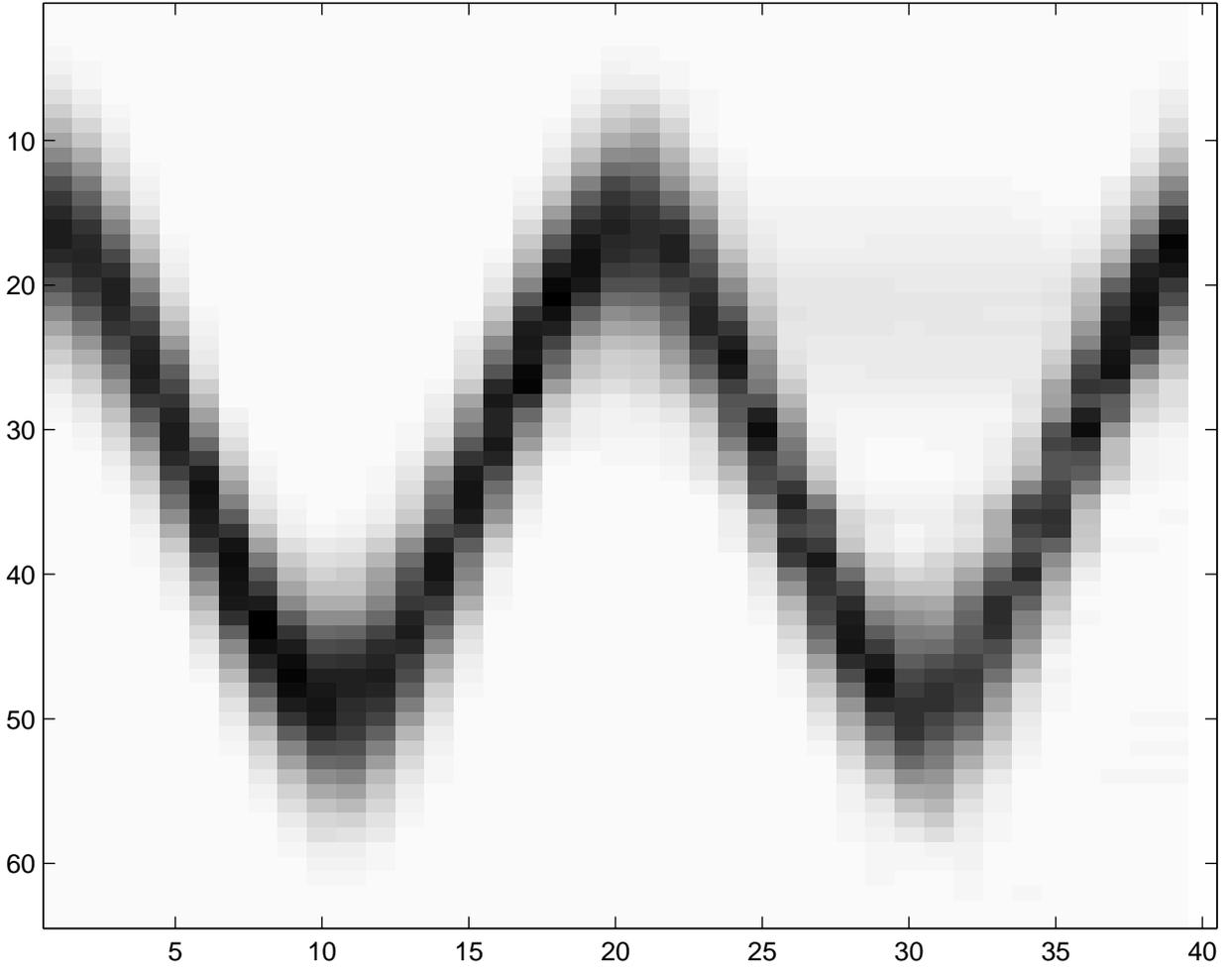
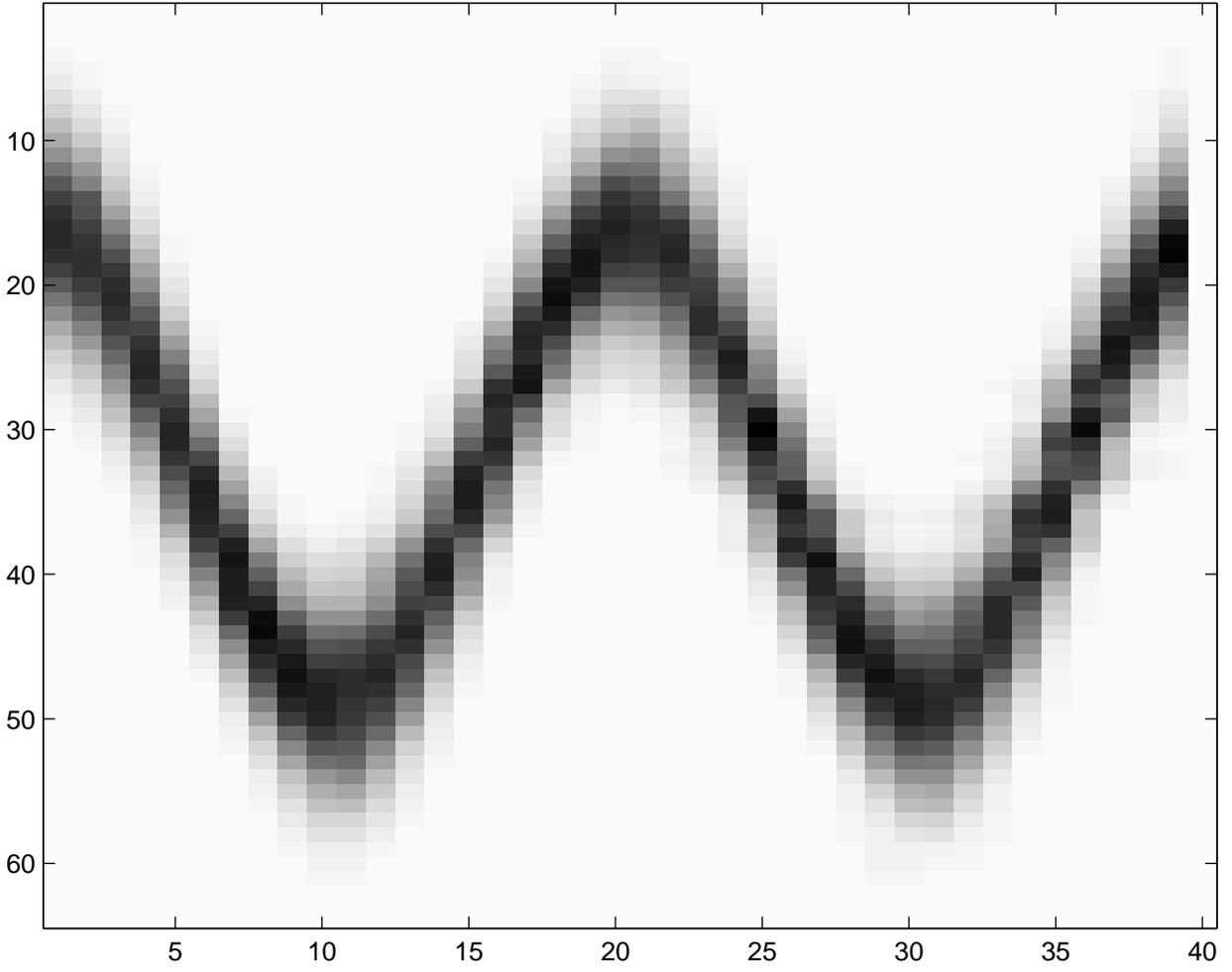
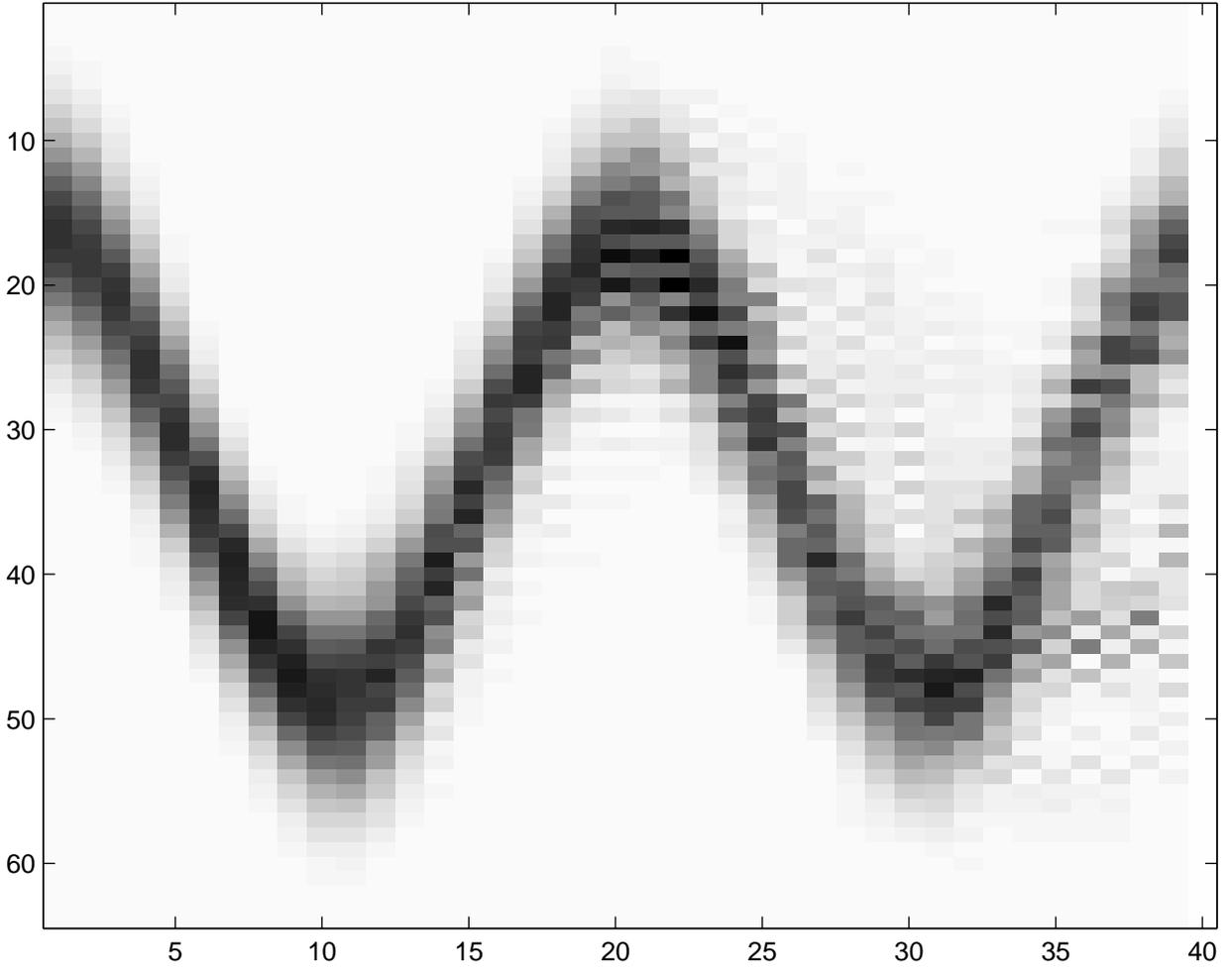
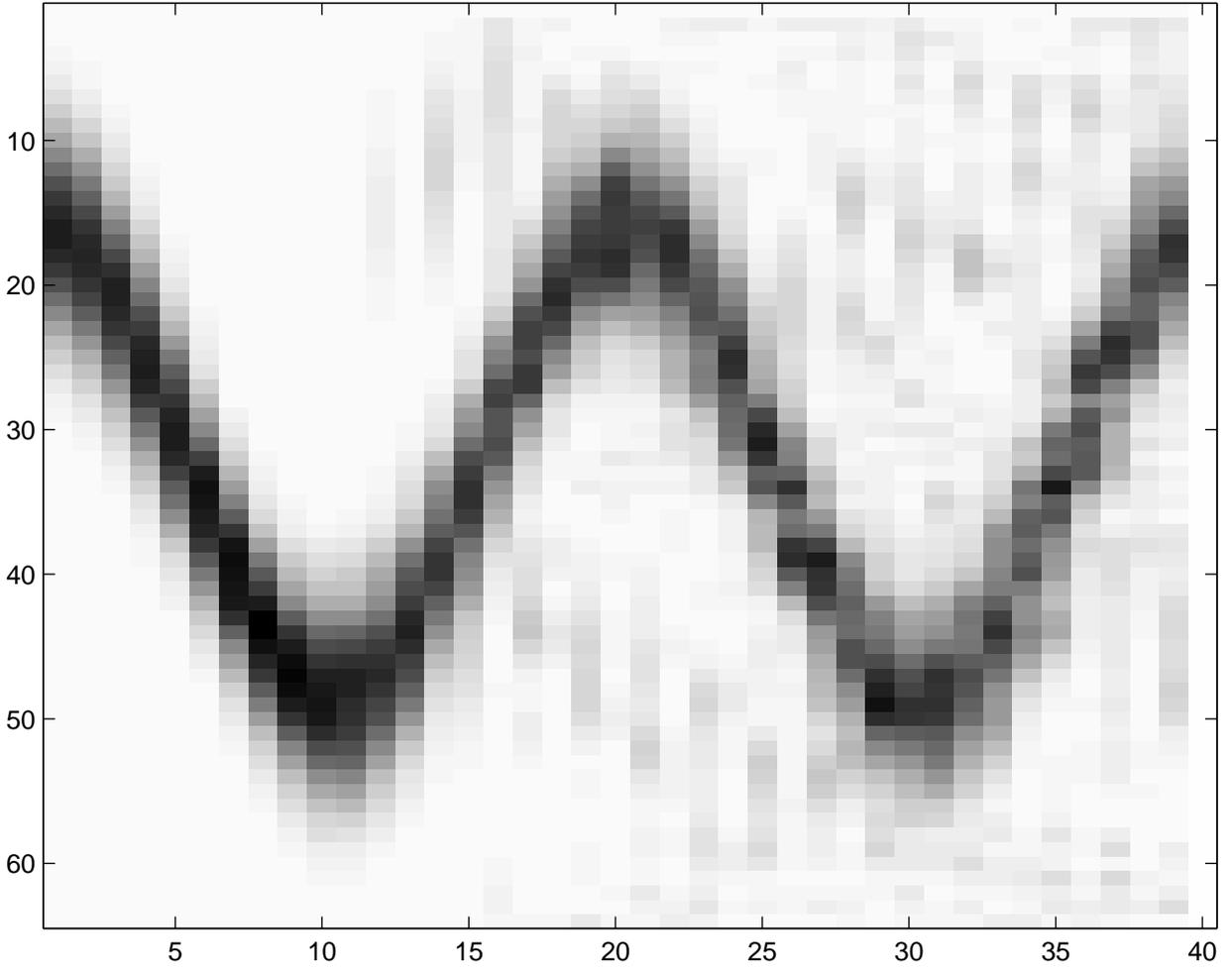
\end{document}